# The Impact of Heatwaves on Population Health: A Large Language Model-Enhanced Agent-Based Simulation

Yuanhao Liu[1], Yuanfei Liu[2], Tian Lu[3], Hengyang Zhang[4], Zuowei Wang[5], and Ying Dai[6]*

[1] Department of Sociology, Johns Hopkins University, Baltimore, MD, USA

[2] Department of Psychiatry, University of Cambridge, Cambridge, UK

[3] Department of Computer Science, Northeastern University, Boston, MA, USA

[4] West China School of Medicine, Sichuan University, Chengdu, China

[5] Department of Journalism and Communication, Zhejiang University, Hangzhou, China

[6] School of Nursing, LKS Faculty of Medicine, The University of Hong Kong, Hong Kong SAR, China

* Corresponding to:

Ying Dai, PhD, RN,

Research Assistant Professor,

School of Nursing, LKS Faculty of Medicine,

The University of Hong Kong

+852 3910 3878 | yindai@hku.hk

**The Impact of Heatwaves on Population Health: A Large Language Model-Enhanced Agent-Based Simulation**

**Abstract:** Extreme heat events, intensified by climate change, threaten public health, yet the socio-behavioral mechanisms of resilience remain unclear. We use a Large Language Model–enhanced agent-based model to simulate a virtual society's response to a prolonged heatwave. One hundred heterogeneous agents receive a Heat Vulnerability Index derived from demographic risk factors and are observed over 13 days spanning baseline, heatwave, and recovery. Preliminary results show primarily psychosocial impacts distributed inequitably: highly vulnerable agents exhibit steeper declines in perceived safety and social connection than low-vulnerability peers. Vulnerability also shapes adaptive capacity: resilient agents sustain routine self-care, whereas the most vulnerable display "behavioral constriction," reducing protective actions. At the collective level, risk-information diffusion follows complex contagion, with adoption driven by social reinforcement in cohesive networks rather than broad exposure. These results demonstrate the value of LLM-enhanced simulation for revealing mechanisms of climate resilience and suggest interventions should leverage community structures beyond mass alerts.

## Introduction

Climate change is escalating the frequency and intensity of extreme heat events, which pose a significant threat to public health, leading to increased mortality, emergency department visits, and hospitalizations in the U.S. annually (CDC, 2023). The negative impacts extend beyond direct health effects to food security (Ray et al., 2019), social conflict (Miles-Novelo & Anderson, 2023), and mental well-being (Pearce et al., 2022), with disproportionate effects on vulnerable populations. While the correlation between heat and adverse outcomes is established, the underlying socio-behavioral mechanisms remain poorly understood (Li at al., 2021). Traditional statistical methods often fail to capture the dynamic, non-linear interactions that shape human adaptation during environmental crises. To address this factor-based limitation on struggling to capture the emergent properties of social systems, AMB is a powerful approach (Macy and Willer 2002).

In addition, modern societal crises, from the COVID-19 pandemic to the escalating impacts of climate change, are profoundly complex phenomena. They unfold across multiple levels, simultaneously affecting individual psychology and behavior, reshaping social networks, and stressing environmental and economic systems. The digital trace data produced in our contemporary world has created a data-rich environment, offering unprecedented insight into mental health, daily activities, and social interactions (Inostroza et al., 2016; Badr et al., 2023).

However, traditional ABMs often rely on simplified behavioral rules, rational-actor assumptions, or stylized mathematical functions that fail to capture the "high-dimensional" realism of human cognition and sociality (Bruch and Atwell 2015). Integrating rich, qualitative data on psychology, culture, and context into these models has proven exceptionally difficult.

To address this gap, we leverage a novel simulation paradigm: a Large Language Model enhanced Agent-Based Model (Park et al. 2023; Anthis et al. 2025; Kozlowski and Evans 2025; Piao et al. 2025). This approach enables the creation of a "high-dimensional" virtual society where heterogeneous agents, endowed with rich psychological and social profiles, interact and adapt in response to a simulated heatwave (Bruch and Atwell 2015; Haase and Pokutta 2025). Our study is grounded in established, cross-level behavioral theories. At the micro-individual level, we model agent needs based on Maslow's Hierarchy. To bridge these micro-agent actions to macro-phenomena, we leverage Complex Contagion Theory to investigate emergent interactive behaviors. In particular, we investigated two primary questions: (*1) How do individuals with different levels of vulnerability emotionally and behaviorally respond to a prolonged heatwave? (2) What social network structures shape the diffusion of risk information and adaptive behaviors?*

By building an artificial society from the ground up, LLM enhanced agent-based simulation provides a computational testbed to explore the micro-to-macro linkages that are often invisible to other methods. This approach allows us to not only move beyond correlation and toward a mechanistic understanding of how individual decisions but also allow as to explore complex empirical contexts influenced by psychological states and environmental cues, aggregate to produce "high dimensional" population-level resilience or vulnerability. In this study, we apply this framework to investigate the multifaceted impacts of heatwaves. We simulate a society of AI agents to explicitly model how environmental stressors propagate through individual psychology, daily behaviors, and social interaction networks, thereby generating observable and testable patterns of population health and social dynamics.

**Methodology: The LLM-Enhanced Agent-Based Model Paradigm**

The recent maturation of Large Language Models (LLMs) offers a transformative solution to the "high dimensional" challenge. By leveraging LLMs as the cognitive engine for each agent, we can move beyond predefined rules and instead endow agents with the capacity for reasoning, memory, and adaptive planning based on their unique profiles and experiences. This LLM-enhanced ABM paradigm makes it possible to instantiate agents with complex, realistic personas derived from real demographic and psychological data, allowing them to perceive, interpret, and react to novel situations—including dynamic environmental stressors and public health messages—in a human-like manner (Park et al. 2023; Anthis et al. 2025; Kozlowski and Evans 2025; Piao et al. 2025).

To implement this approach, we utilize AgentSociety, a large-scale simulation platform designed specifically for LLM-driven generative social science. The platform integrates three core components: (1) LLM-driven generative agents with psychologically grounded minds, (2) a realistic societal environment that provides dynamic feedback, and (3) a powerful engine enabling large-scale simulations. The theoretical design of AgentSociety is rooted in established social science theories; agent needs are structured according to Maslow's hierarchy of needs, while their behavioral decision-making is modeled on the theory of planned behavior (Piao et al., 2025).

While the LLM-enhanced ABM paradigm represents a significant leap forward, scholars have pointed several limitations that define the frontier for future work. Among them, most importantly, while LLM-based agents produce highly realistic and plausible behaviors, these are ultimately simulations of human action, not perfect replications (Kozlowski and Evans 2025). The ultimate test of this paradigm lies in rigorous empirical validation. As outlined in our discussion, the next phase of this research will involve directly comparing the model's predictions against observational, real-world data on behavioral and health outcomes during heatwaves to systematically calibrate the model and assess its external validity.

**Simulation Model**

For this study, we populated the preliminary simulation with 100 heterogeneous agents and used the Qwen-plus LLM to power their cognitive functions (Bai et al. 2025). We instantiated a virtual society of 100 generative agents with heterogeneous demographic profiles derived from real-world data. To quantify individual susceptibility, we constructed a Heat Vulnerability Index (HVI) based on literature-defined risk factors, including age, socioeconomic status, race, and residential characteristics. Agents were classified into four vulnerability levels: Low, Moderate, High, and Extreme (Cheng et al., 2021; Li et al., 2024; Inostroza et al., 2016). See agent demographic and HVI details in Appendix A.

The simulation unfolded over 13 days, divided into three phases: a 3-day Baseline, a 5-day Heatwave (40°C), and a 5-day Recovery period. Government broadcasts announcing the heatwave's start and end were disseminated to all agents to simulate public information campaigns. Throughout the simulation, we collected time-stamped data on agents' emotional states, behaviors, satisfaction of physiological and social needs, and interpersonal communications.

Our analytical strategy is designed to examine the heatwave's impact across multiple social scales. First, at the micro-individual level, we track the evolution of agents' emotional states (e.g., fear, joy) and daily actions (e.g., self-care, work) to establish the direct effects of environmental stress. Second, at the meso-group level, we use the HVI to stratify our analysis, employing mixed-effects models to test for differential impacts on the psychological and behavioral trajectories of vulnerable versus resilient populations (Piao et al., 2025). Finally, at the macro-collective level, we analyze the communication networks between agents to investigate the social diffusion of risk information, testing competing theories of simple versus complex contagion to understand emergent, population-wide responses (Centola and Macy 2007). This multi-level approach allows us to build a comprehensive, mechanistic account of social resilience to climate shocks.

## Preliminary Results

Our preliminary analysis reveals three key findings regarding the differential impacts of the simulated heatwave, the nature of behavioral adaptation, and the social mechanisms of information diffusion. (Descriptive results detailing the overall temporal trends in agent emotions and actions are provided in Appendix B).

### 1. Disproportionate Psychosocial Impacts on Vulnerable Populations

The heatwave did not affect all agents equally. Table 1 reports two-level mixed-effects models of psychosocial preliminary results (See model details in Appendix C), we found that agents' psychosocial needs were far more sensitive to the environmental stressor than their physiological needs. While changes in hunger or tiredness needs did not significantly differ across vulnerability groups, the satisfaction of safety and social needs degraded more rapidly for vulnerable individuals. The interaction between simulation day and HVI class was highly significant, revealing that agents in the "Extreme" vulnerability class experienced a steeper decline in their sense of safety over time ($\beta$ = -0.03/day, p = 0.001) compared to the "Low" vulnerability group. Similarly, social need satisfaction eroded most quickly among highly vulnerable agents, underscoring that the primary impact of the heatwave manifested as a psychosocial, rather than purely physiological, crisis for those most at risk.

**Table 1. Two-level mixed-effects modeling of health vulnerability and daily needs satisfaction**

| | **Hungry** | | | **Safe** | | | **Social** | | | **Tired** | | |
|---|---|---|---|---|---|---|---|---|---|---|---|---|
| *Predictors* | *Estimates* | *CI* | *p* | *Estimates* | *CI* | *p* | *Estimates* | *CI* | *p* | *Estimates* | *CI* | *p* |
| (Intercept) | 0.38 | 0.35 – 0.40 | **<0.001** | 0.30 | 0.10 – 0.50 | **0.003** | -0.08 | -0.30 – 0.14 | 0.479 | 0.22 | 0.20 – 0.23 | **<0.001** |
| day | 0.00 | -0.00 – 0.00 | 0.106 | 0.04 | 0.03 – 0.04 | **<0.001** | 0.05 | 0.05 – 0.06 | **<0.001** | -0.00 | -0.01 – -0.00 | **<0.001** |
| HVI class [Moderate] | 0.01 | -0.03 – 0.05 | 0.635 | 0.04 | -0.23 – 0.30 | 0.783 | 0.14 | -0.14 – 0.43 | 0.326 | 0.01 | -0.01 – 0.03 | 0.235 |
| HVI class [High] | 0.01 | -0.04 – 0.05 | 0.737 | 0.08 | -0.23 – 0.39 | 0.614 | 0.14 | -0.21 – 0.48 | 0.436 | -0.01 | -0.03 – 0.01 | 0.339 |
| HVI class [Extreme] | 0.00 | -0.06 – 0.06 | 0.912 | 0.00 | -0.43 – 0.44 | 0.986 | 0.13 | -0.35 – 0.61 | 0.591 | 0.02 | -0.01 – 0.05 | 0.297 |
| day × HVI class [Moderate] | -0.00 | -0.01 – -0.00 | **0.012** | -0.04 | -0.05 – -0.03 | **<0.001** | -0.06 | -0.06 – -0.05 | **<0.001** | 0.00 | -0.00 – 0.00 | 0.941 |
| day × HVI class [High] | -0.00 | -0.01 – -0.00 | **0.021** | -0.04 | -0.06 – -0.03 | **<0.001** | -0.06 | -0.06 – -0.05 | **<0.001** | 0.00 | -0.00 – 0.00 | 0.515 |
| day × HVI class [Extreme] | -0.00 | -0.01 – 0.00 | 0.114 | -0.03 | -0.05 – -0.01 | **0.001** | -0.06 | -0.07 – -0.05 | **<0.001** | -0.00 | -0.00 – -0.00 | **0.042** |
| **Random Effects** | | | | | | | | | | | | |
| $\sigma^2$ | 0.43 | | | 3.85 | | | 1.26 | | | 0.07 | | |
| $\tau_{00}$ | 0.00 $_{id}$ | | | 0.29 $_{id}$ | | | 0.37 $_{id}$ | | | 0.00 $_{id}$ | | |
| ICC | 0.01 | | | 0.07 | | | 0.23 | | | 0.01 | | |
| N | 100 $_{id}$ | | | 100 $_{id}$ | | | 100 $_{id}$ | | | 100 $_{id}$ | | |
| Observations | 61200 | | | 61200 | | | 61200 | | | 61200 | | |
| Conditional $R^2$ | 0.009 | | | 0.072 | | | 0.235 | | | 0.019 | | |

We also report emotional impact in Appendix D

## 2. Vulnerability Determines Adaptive Behavioral Strategies

Agents' behavioral responses were systematically shaped by their vulnerability status, revealing distinct patterns of adaptive capacity. As shown in Figure 1, low and moderate-vulnerability agents primarily maintained high levels of routine self-care, suggesting a capacity to absorb the stressor without fundamentally altering their daily priorities. In contrast, high-vulnerability agents executed a strategic shift, dramatically increasing heatwave-specific responses (e.g., hydrating, using A/C) while still maintaining self-care, indicating an active but strained adaptation. Most strikingly, agents in the "Extreme" vulnerability class exhibited behavioral constriction, with markedly lower engagement across all action categories, including self-care and heatwave responses. This suggests that extreme vulnerability may overwhelm adaptive capacity, leading to a dangerous reduction in protective behaviors when they are needed most.

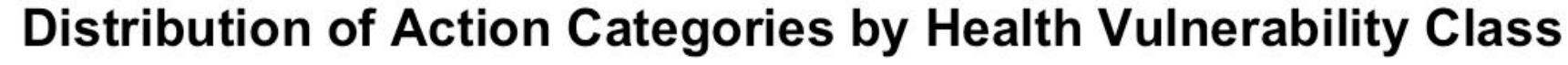

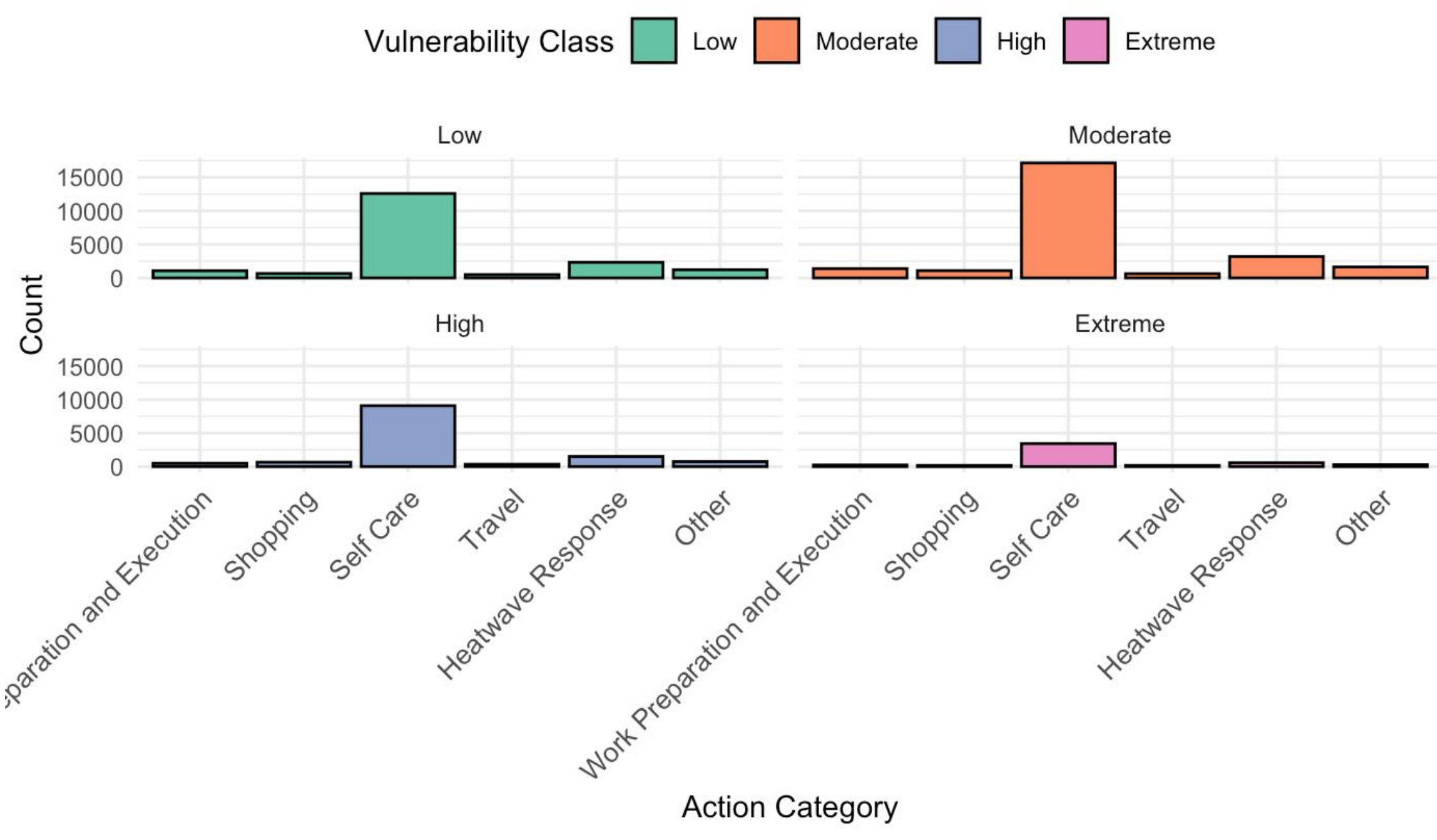

Figure 1. Agent behaviors across health vulnerability groups

## 3. Social Diffusion of Risk Information Follows Complex Contagion

To understand how agents collectively responded to the crisis, we analyzed the diffusion of heatwave-related discussion within their social networks (See model details in Appendix E). Our logistic regression analysis tested whether an agent would begin discussing the heatwave as a function of their network position on the previous day. The results in Table 2 provide strong support for the theory of complex contagion (Centola and Macy 2007; Guilbeault et al., 2018). The likelihood of an agent adopting the topic was significantly predicted by the number of friends who were already discussing it (informed friends, $\beta = 0.411$, $p < 0.001$) and the cohesiveness of their local network (clustering coefficient, $\beta = 0.903$, $p < 0.001$). Conversely, an agent's total number of connections (degree) was negatively associated with adoption ($\beta = -0.025$, $p < 0.001$), suggesting that simple, broad exposure is less effective than social reinforcement from a close-knit group. This finding has critical implications, suggesting that public health messages are most likely to gain traction and translate into collective awareness within cohesive community structures.

**Table 2. Logistic Regression Predicting Heatwave Expression (Lagged Network Features)**

| Variable | Coefficient | Std. Error | z | p-value | 95% CI (Lower) | 95% CI (Upper) |
|---|---|---|---|---|---|---|
| Constant | -1.7523 | 0.034 | -51.80 | 0.000 | -1.819 | -1.686 |
| Degree (t–1) | -0.0254 | 0.006 | -4.55 | 0.000 | -0.036 | -0.014 |
| Clustering Coefficient (t–1) | 0.9031 | 0.093 | 9.72 | 0.000 | 0.721 | 1.085 |
| Informed Friends (t–1) | 0.4111 | 0.008 | 52.17 | 0.000 | 0.396 | 0.427 |

*N = 38,064 user-day observations. Pseudo $R^2$ = 0.0835. All coefficients estimated via logistic regression.*

## Discussion

In this study, we simulated an agent-based society encountering a heatwave to evaluate the social mechanisms that shape population health and resilience during an extreme weather event. Our findings indicate that agents' psychosocial needs are more sensitive to environmental changes than their physiological needs, that vulnerability status systematically determines adaptive behavioral strategies, and that the social diffusion of risk information follows patterns of complex contagion. These results offer both methodological and empirical insights for population studies and climate adaptation.

Methodologically, our work demonstrates that LLM-based simulations can reproduce and add nuance to classic findings from computational social science. For instance, our analysis of agent communication networks not only confirms the importance of social contagion but also allows us to adjudicate between competing theories, finding strong support for the role of cohesive, reinforcing ties (complex contagion) over simple exposure in the diffusion of risk discourse. This aligns with foundational work on the spread of health behaviors and provides a mechanistic explanation for why some public health messages succeed while others fail.

Empirically, our findings highlight that the primary threat of a heatwave, at least in a society with stable access to food and shelter, may be psychosocial. The sharp decline in safety and social need satisfaction among vulnerable agents suggests that the psychological stress and social isolation caused by the crisis are the most immediate and inequitably distributed burdens. This resonates with qualitative accounts from real-world disasters, such as the 1995 Chicago heatwave, where social isolation was a key predictor of mortality (Klinenberg 2022). Furthermore, the "behavioral constriction" observed in our most vulnerable agents—whereby their capacity for even basic self-care and protective actions was diminished—points to a critical threshold where environmental stress overwhelms an individual's adaptive capacity.

The policy implications of these findings are significant. The success of complex contagion suggests that top-down, broad-based heatwave alerts may be necessary but insufficient. To be effective, particularly for reaching isolated or marginalized individuals, these messages must be amplified and reinforced through trusted, cohesive community networks. Our results provide a simulated, mechanistic basis for prioritizing community-level interventions that strengthen social ties as a core component of climate resilience strategy.

### Limitations and Empirical Validation (Next Step)

The promise of this new simulation paradigm is accompanied by important limitations that define our path forward. The primary limitation is the inherent gap between a simulated reality and the empirical world; while our agents produce plausible, theoretically grounded behaviors, they are not perfect human replicas. The current model's simplifications—such as its constrained population size and the absence of a detailed physiological health module—also limit its generalizability.

Addressing these limitations requires rigorous empirical validation (Kozlowski and Evans 2025), which constitutes the next phase of our research agenda. Our plan involves a multi-pronged approach to directly compare the simulation's outputs against observational data from real-world heatwave events.

1. Behavioral and Social Validation: We will source geolocated social media data to track the real-world diffusion of heatwave discourse and anonymized human mobility data to measure changes in activity patterns. These empirical benchmarks will allow us to test and calibrate our model's predictions regarding complex contagion and behavioral adaptation.

2. Health Outcome Validation: We will use public health surveillance data, such as records of emergency department visits for heat-related illness, stratified by geographic areas with varying levels of socioeconomic vulnerability. By comparing these empirical health outcomes to the distress patterns of our simulated HVI groups, we can assess the model's external validity in predicting health inequities.

Through this process of pattern-oriented validation and calibration, we will systematically refine the agent parameters to close the gap between the simulation and reality. By grounding our artificial society in empirical data, we aim to develop this platform into a reliable computational testbed for evaluating the effectiveness of different public health interventions and designing more equitable policies to enhance social resilience in the face of a changing climate.

## Appendix A. Agent Setup

The experiment is structured into several phases: (1) Baseline Phase: Three days of normal weather establish behavioral and psychological baselines for all agents. (2) Initial Assessment: On day 2, an official broadcast warning (Broadcast1) delivered. (3) Heatwave Intervention: On day 4, the environment is programmatically altered to simulate a severe heatwave (40°C). This phase lasts five days, during which agents' needs, behaviors, and psychological states are closely monitored. (4) Recovery Phase: Environmental parameters are restored to normal. After one day, a subsequent broadcast (Broadcast2) signals the end of the heatwave, followed by four additional days of observation.

A population of 100 citizen agents is instantiated, each with a heterogeneous profile generated from real demographic data and randomized behavioral properties using a custom memory configuration function. Each agent's memory includes attributes such as age, occupation, income, satisfaction of core needs (hunger, safety, social), emotional state, skills, and social relationships. This ensures diversity and realism in the simulated responses. We generated a total of 100 resident agents with a mean age of 55.1 years (SD = 19.5). Residential distribution was balanced among urban (37%), suburban (35%), and rural (28%) areas. Detailed demographic characteristics were presented in Appendix A Table 1.

**Appendix A Table 1. Demographic characteristics of generated resident agents**

| | **Female (N=53)** | **Male (N=47)** | **Overall (N=100)** |
|---|---|---|---|
| Age | | | |
| Mean (SD) | 55.7 (20.4) | 54.3 (18.8) | 55.1 (19.5) |
| Median [Min, Max] | 58.0 [18.0, 85.0] | 53.0 [19.0, 84.0] | 57.0 [18.0, 85.0] |
| Race | | | |
| Asian | 2 (3.8%) | 1 (2.1%) | 3 (3.0%) |
| Black | 17 (32.1%) | 19 (40.4%) | 36 (36.0%) |
| Other | 4 (7.5%) | 3 (6.4%) | 7 (7.0%) |
| White | 30 (56.6%) | 24 (51.1%) | 54 (54.0%) |
| Education attainment | | | |
| Bachelor and above | 25 (47.2%) | 23 (48.9%) | 48 (48.0%) |
| Below bachelor | 28 (52.8%) | 24 (51.1%) | 52 (52.0%) |
| Occupation | | | |
| Artist | 3 (5.7%) | 7 (14.9%) | 10 (10.0%) |
| Athlete | 6 (11.3%) | 3 (6.4%) | 9 (9.0%) |
| Businessman | 8 (15.1%) | 4 (8.5%) | 12 (12.0%) |
| Doctor | 3 (5.7%) | 9 (19.1%) | 12 (12.0%) |
| Engineer | 7 (13.2%) | 3 (6.4%) | 10 (10.0%) |
| Manager | 9 (17.0%) | 7 (14.9%) | 16 (16.0%) |
| Other | 8 (15.1%) | 5 (10.6%) | 13 (13.0%) |
| Student | 6 (11.3%) | 5 (10.6%) | 11 (11.0%) |
| Teacher | 3 (5.7%) | 4 (8.5%) | 7 (7.0%) |
| Skill | | | |
| Good at communication | 9 (17.0%) | 10 (21.3%) | 19 (19.0%) |
| Good at creativity | 11 (20.8%) | 6 (12.8%) | 17 (17.0%) |
| Good at problem-solving | 12 (22.6%) | 7 (14.9%) | 19 (19.0%) |
| Good at teamwork | 12 (22.6%) | 11 (23.4%) | 23 (23.0%) |
| Other | 9 (17.0%) | 13 (27.7%) | 22 (22.0%) |
| Consumption | | | |
| High | 18 (34.0%) | 11 (23.4%) | 29 (29.0%) |
| Low | 6 (11.3%) | 4 (8.5%) | 10 (10.0%) |
| Median | 9 (17.0%) | 14 (29.8%) | 23 (23.0%) |
| Slightly high | 9 (17.0%) | 9 (19.1%) | 18 (18.0%) |
| Slightly low | 11 (20.8%) | 9 (19.1%) | 20 (20.0%) |
| Income | | | |
| Mean (SD) | 81100 (91100) | 56100 (37800) | 69400 (71900) |
| Median [Min, Max] | 49600 [8140, 500000] | 46800 [7640, 194000] | 49300 [7640, 500000] |
| Residence | | | |
| City | 17 (32.1%) | 20 (42.6%) | 37 (37.0%) |
| Rural | 17 (32.1%) | 11 (23.4%) | 28 (28.0%) |
| Suburb | 19 (35.8%) | 16 (34.0%) | 35 (35.0%) |
| Personality | | | |
| Ambivert | 9 (17.0%) | 9 (19.1%) | 18 (18.0%) |
| Extrovert | 12 (22.6%) | 13 (27.7%) | 25 (25.0%) |
| Introvert | 17 (32.1%) | 13 (27.7%) | 30 (30.0%) |
| Outgoint | 15 (28.3%) | 12 (25.5%) | 27 (27.0%) |
| Religion | | | |
| Buddhist | 5 (9.4%) | 10 (21.3%) | 15 (15.0%) |
| Christian | 15 (28.3%) | 5 (10.6%) | 20 (20.0%) |
| Hindu | 9 (17.0%) | 6 (12.8%) | 15 (15.0%) |
| Muslim | 8 (15.1%) | 10 (21.3%) | 18 (18.0%) |
| None | 9 (17.0%) | 6 (12.8%) | 15 (15.0%) |
| Other | 7 (13.2%) | 10 (21.3%) | 17 (17.0%) |
| Marital status | | | |
| Divorced | 12 (22.6%) | 15 (31.9%) | 27 (27.0%) |
| Married | 14 (26.4%) | 12 (25.5%) | 26 (26.0%) |
| Not married | 8 (15.1%) | 9 (19.1%) | 17 (17.0%) |
| Widowed | 19 (35.8%) | 11 (23.4%) | 30 (30.0%) |
| Health vulnerability index | | | |
| Low | 13 (24.5%) | 17 (36.2%) | 30 (30.0%) |
| Moderate | 25 (47.2%) | 16 (34.0%) | 41 (41.0%) |
| High | 11 (20.8%) | 10 (21.3%) | 21 (21.0%) |
| Extreme | 4 (7.5%) | 4 (8.5%) | 8 (8.0%) |

We generated the heat vulnerability index (HVI) through a systematic approach that integrated demographic and socioeconomic factors known to influence heat-related health risks. First, key vulnerability indicators were selected based on literature review, including age (with heightened vulnerability for those under 18 or over 65) (Cheng et al., 2021; Inostroza et al., 2016), socioeconomic status (combining income and education levels), race/ethnicity (accounting for documented disparities) (Inostroza et al., 2016), residential characteristics (Li et al., 2024), and health proxies (using marital status as an indicator of social isolation) (Li et al., 2024). These variables were normalized and coded on a standardized scale (1-3) to ensure comparability, with higher values indicating greater vulnerability. An expert-weighted method was used to assign predetermined weights based on theoretical importance (0.25 for age, 0.30 for socioeconomic factors, 0.15 for race/ethnicity, 0.2 for residence area, and 0.1 for health proxies). The final HVI score was calculated as a composite of these weighted components, with results classified into four vulnerability levels (Low, Moderate, High, and Extreme) using percentile-based thresholds. The HVI distribution was presented in Appendix A Figure 1.

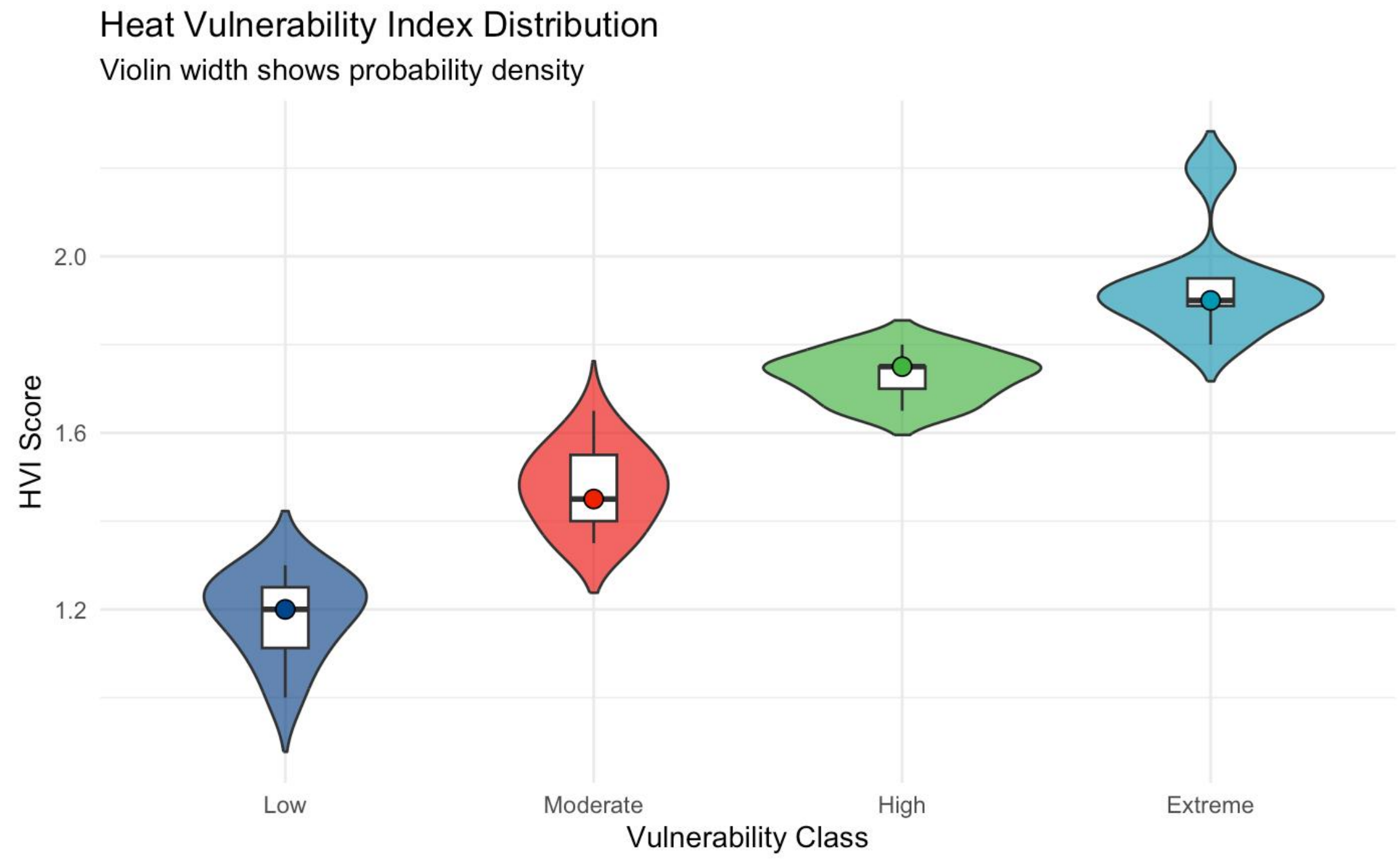

Appendix A Figure 1. Agents’ health vulnerability index (HVI)

## Appendix B. Simulation Descriptive Results

Throughout the simulation, agent needs satisfaction metrics (hunger, safety, social) are periodically extracted and stored for analysis via a custom metric extractor at regular intervals. Simulation outputs, including agent status, dialogue history and survey responses, are automatically generated. After the experiment, we make downstream statistical analysis and visualization based on simulation records. Specifically, the following data were extracted: time stamps, steps, agent id, demographic information such as age, gender, race/ethnicity, etc., agent Maslow's hierarchy needs, including hunger needs, safety needs, tiredness needs, and social needs, with each need ranges between 0-1 and higher scores indicating higher needs priority, six types of emotion status (including sadness, joy, fear, disgust, anger, and surprise), each emotion status ranges between 0-10, with higher scores indicating higher intensity, agent behaviors, and agent texting messages across time.

Appendix B Figure 1 report agent emotional status across time, which reveals significant phase-dependent variations. During baseline periods, gratification (38.2%) and joy (13.8%) were predominant, while the notification of an impending heatwave at Day 2 triggered the emergence of fear (0.9%), distress (0.1%), alertness (0.1%), and concern (1.7%). Notably, relief exhibited a sharp decline from 38.2% on Day 1 to 2.1% on Day 2, remaining at low levels throughout the heatwave period before showing a minor rebound during recovery. The recovery phase was characterized by three distinct emotional patterns: sustained gratification (32.2 - 43.3%), a resurgence of joy (32.0 - 36.3%), and the near-absence of relief (< 0.7% after Day 8). Fear manifestations displayed temporal clustering during thermal stress events (Days 3 - 7), while satisfaction progressively declined from baseline (26.9%) to recovery (15.5%). These findings underscore the dynamic responsiveness of emotional states to environmental stressors and subsequent recovery periods.

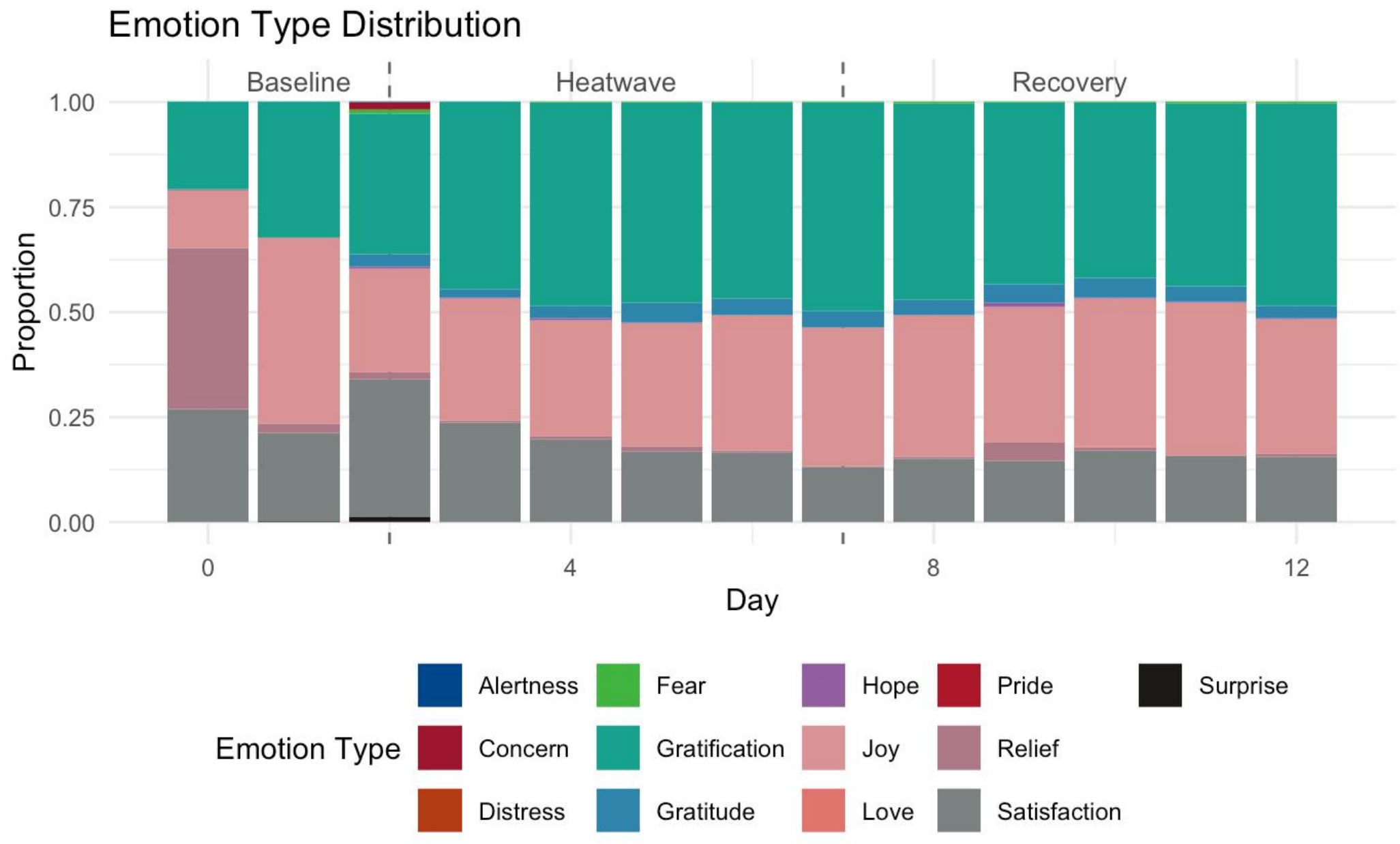

Appendix B Figure 1. Agents' emotion type distribution across time

Appendix B Figure 2 reports agent behavior trend across time. The agents showed distinct behavioral patterns across three experimental phases (Figure 3). During the baseline period (Days 0 - 2), self-care actions dominated (Mean ≈ 3,500 counts/day), while work-related activities (preparation/execution) showed an initial presence (Mean ≈ 900 counts/day) followed by a gradual decline. The heatwave phase (Days 3 - 7) precipitated significant behavioral adaptation, marked by the emergence of heatwave-specific responses such as using air conditioning or a fan, reducing sunlight exposure, and hydrating with water or electrolyte-rich drinks (peak ≈ 1,000 counts) alongside sustained self-care maintenance (Mean ≈ 3,500 counts/day). Work-related actions demonstrated continued attenuation (Δ ≈ -40% from baseline), suggesting thermal stress-induced productivity modulation.

The recovery phase (Days 8-12) showed partial behavioral normalization, with self-care actions maintaining elevated levels (≈3,500 counts/day) and heatwave responses declining to near-baseline levels (Δ ≈ -80% from peak). Notably, shopping and travel activities remained stable throughout all phases (±15% variation), indicating these behaviors were relatively unaffected by thermal conditions. The phase transitions (Days 2 and 7) were marked by abrupt changes in work-related actions and heatwave responses, suggesting threshold effects in behavioral adaptation.

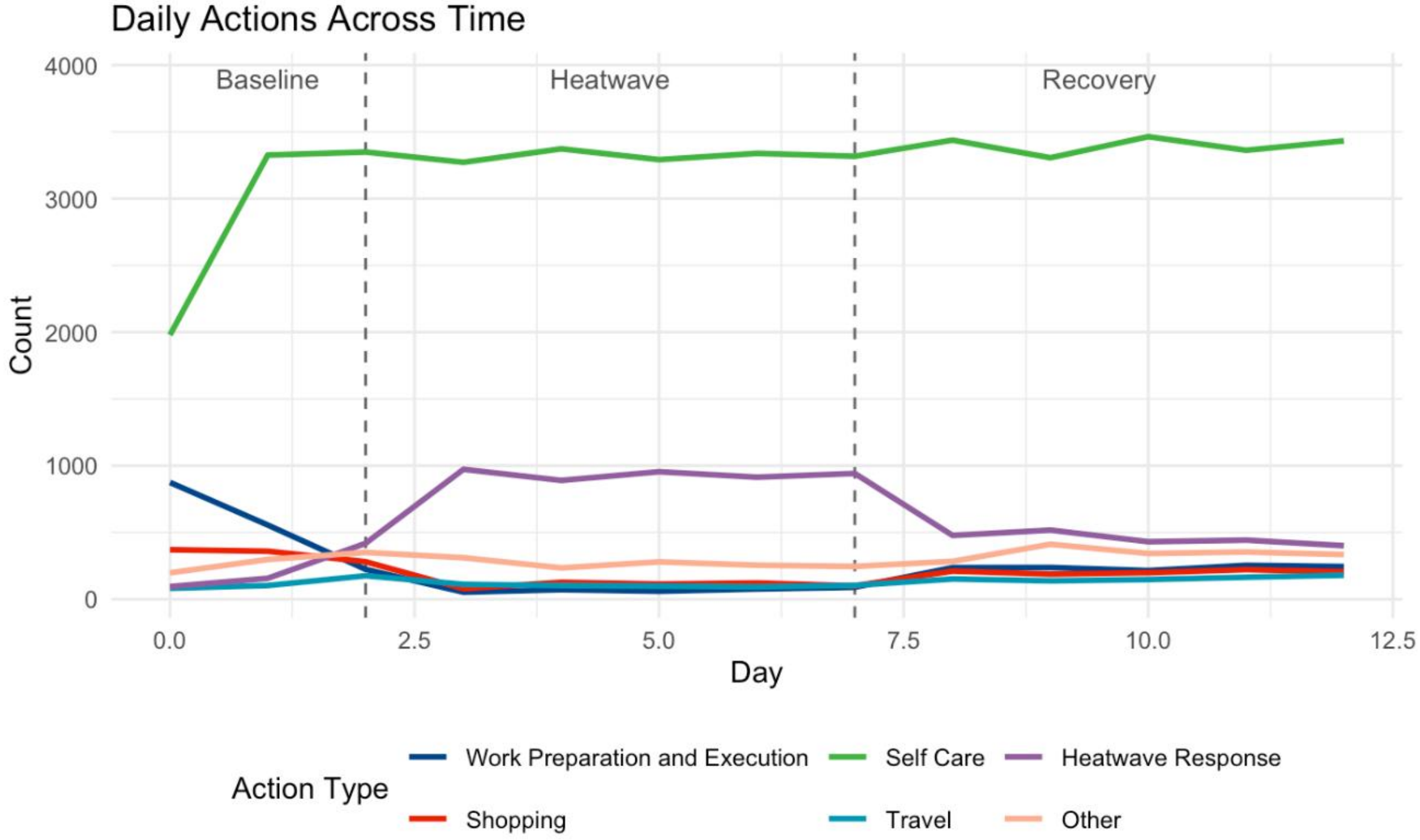

Appendix B Figure 2. Agents' daily actions across time

**Appendix C. Simulation Two-level mixed-effects modeling on Needs of Vulnerable Groups**

Descriptive statistics (e.g., means, medians, standard deviations, frequencies, and proportions) were used to describe agent demographic characteristics, as appropriate. Linear regression was used to examine the associations between demographic variables and the perception of heat waves among these agents. Two-level mixed-effects modeling was conducted to explore the pattern of agents' physiological (hungry and safety), emotional needs and behaviors, and mental health across simulation days. Simulation day was entered as the level 1 variable, whereas predictors (agent physiological needs, emotional needs, mental health) and other covariates were included as level 2 variables. The interaction term "day*$Uij$" was used to test whether agent physiological and emotional trajectories were significantly different between health vulnerability groups.

The composite model can be written as: $Yij = \beta 0 + \beta 1(\text{day}ij) + \beta 2(Uij) + \beta 3(\text{day}ij * Uij) + \alpha i + \gamma i(\text{day}ij) + \varepsilon ij$, where Yij refers to agent physiological or emotional outcomes for agent i at day j. Day refers to simulation days, U refers to health vulnerability class (Low, Moderate, High, and Extreme), and "day * U" refers to the interaction between simulation day and health vulnerability class. β0 is the global intercept, and β1 and β2 are the fixed effects for simulation day and the explanatory variable/covariate, respectively. αi, γi, and εij are the random effects of the composite model. αi and γi follow the multivariate normal distribution. The interaction term (β3) was used to test whether agent physiological and emotional trajectories were significantly different between the health vulnerability classes.

Mixed-effects linear modeling with restricted maximum likelihood estimation (REML) was conducted with the "lmerTest" package. All analyses were conducted with R (version 4.4.0), with a $p < 0.05$ representing statistical significance. All code, configuration files, and workflow definitions used in this study are open-sourced and can be found at: https://github.com/Xander-run/AgentSociety/tree/experiment-1.3.7/examples/heatwave

The two-level mixed-effects models revealed significant temporal dynamics in daily needs satisfaction across health vulnerability classes (Table 2 and Figure 4). Significant day × HVI class interactions revealed that higher vulnerability groups experienced steeper declines in safety (Extreme: $\beta = -0.03$/day, $p = 0.001$) and social satisfaction (High: $\beta = -0.06$/day, $p < 0.001$) over time compared to low-vulnerability counterparts. Notably, hunger and tiredness showed minimal vulnerability-based differences (all $p > 0.05$). The random effects structure indicated substantial between-individual variability in social needs (ICC = 0.23) compared to other domains (ICCs 0.01 - 0.07), suggesting greater personalization in social satisfaction trajectories. These patterns remained robust across 61,200 observations, with model fit statistics (conditional $R^2$) ranging from 0.009 (hunger) to 0.235 (social), indicating the models explained up to 23.5% of variance in social satisfaction. Agents with low HVI had significantly higher satisfaction scores in safety needs and social needs when the heatwave finished on Day 8, while agents with extremely high HVI had significantly lower satisfaction scores in tiredness needs (Appendix C Figure 1).

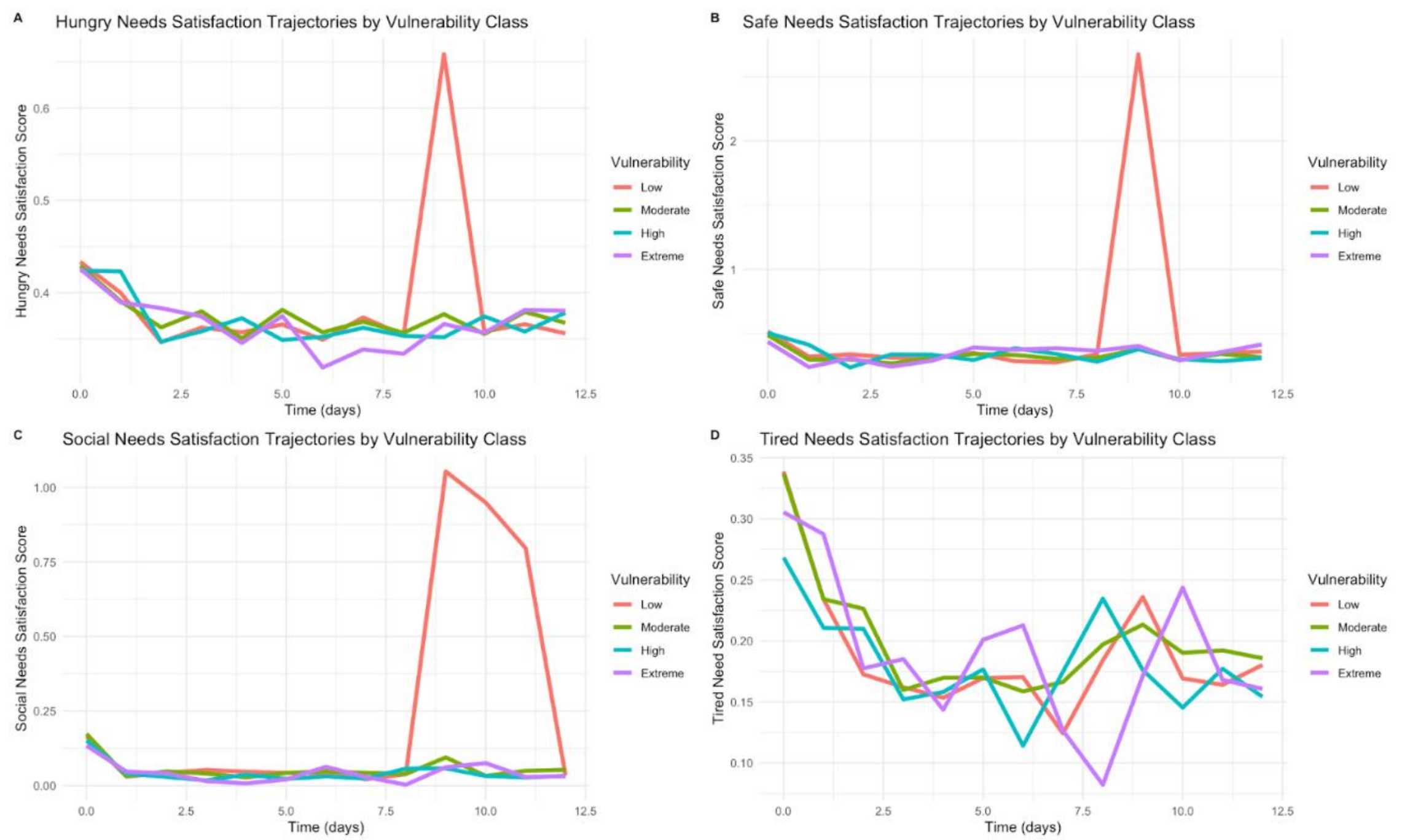

Appendix C Figure 1. Daily needs satisfaction trajectories across HVI groups.

((A) Hunger needs satisfaction; (B) Safety needs satisfaction; (C) Social needs satisfaction; (D) Tiredness needs satisfaction.)

**Appendix D. Simulation Two-level mixed-effects modeling on Emotional Impact**

The longitudinal analysis of emotional trajectories across health vulnerability classes (Low, Moderate, High, Extreme) revealed three key patterns (Appendix D Table 1). First, temporal dynamics dominated emotional changes, with surprise ($\beta = 0.11$/day, $p < 0.001$) and joy ($\beta = 0.10$/day, $p < 0.001$) increasing over time, while negative emotions like anger ($\beta = -0.09$/day) and sadness ($\beta = -0.09$/day) declined (all $p < 0.001$). Second, vulnerability-class interactions showed differential trajectories: Extreme-class individuals exhibited attenuated surprise gains (day×Extreme: $\beta = -0.03$, $p < 0.001$) but accelerated fear reduction (day×Extreme: $\beta = -0.05$, $p < 0.001$) compared to low-vulnerability counterparts. Third, baseline differences emerged in intercepts, with joy highest overall (intercept = 7.77, SE = 0.11) and fear lowest (intercept = 2.17, SE = 0.10), though vulnerability classes showed no baseline differences (all HVI class $p > 0.05$). Random effects indicated substantial between-individual variability (ICCs 0.21-0.50), particularly for surprise (ICC = 0.50) and disgust (ICC = 0.36). Model fits were strongest for surprise ($R^2 = 0.522$) and weakest for anger ($R^2 = 0.347$), suggesting greater predictability of positive/novelty emotions than negative states. These patterns held across 61,200 observations from 100 participants, demonstrating robust emotional adaptation dynamics modulated by vulnerability status.

**Appendix D Table 1. Two-level mixed-effects modeling of health vulnerability and emotional scores**

| | **Surprise** | | **Joy** | | **Fear** | | **Anger** | | **Sadness** | | **Disgust** | |
|---|---|---|---|---|---|---|---|---|---|---|---|---|
| *Predictors* | *Estimates* | *std. Error* | *Estimates* | *std. Error* | *Estimates* | *std. Error* | *Estimates* | *std. Error* | *Estimates* | *std. Error* | *Estimates* | *std. Error* |
| (Intercept) | 4.72 *** | 0.24 | 7.77 *** | 0.11 | 2.17 *** | 0.10 | 1.04 *** | 0.08 | 1.76 *** | 0.07 | 1.51 *** | 0.10 |
| day | 0.11 *** | 0.00 | 0.10 *** | 0.00 | -0.05 *** | 0.00 | -0.09 *** | 0.00 | -0.09 *** | 0.00 | -0.08 *** | 0.00 |
| HVI class [Moderate] | 0.23 | 0.31 | 0.20 | 0.15 | -0.12 | 0.13 | 0.10 | 0.10 | 0.04 | 0.09 | 0.05 | 0.13 |
| HVI class [High] | -0.05 | 0.37 | 0.19 | 0.18 | 0.10 | 0.15 | 0.22 | 0.12 | 0.04 | 0.10 | -0.02 | 0.16 |
| HVI class [Extreme] | -0.34 | 0.52 | 0.14 | 0.25 | -0.27 | 0.21 | 0.01 | 0.17 | -0.07 | 0.14 | -0.18 | 0.22 |
| day × HVI class [Moderate] | -0.05 *** | 0.00 | -0.02 *** | 0.00 | -0.01 *** | 0.00 | -0.02 *** | 0.00 | -0.04 *** | 0.00 | -0.02 *** | 0.00 |
| day × HVI class [High] | -0.03 *** | 0.00 | -0.00 | 0.00 | -0.02 *** | 0.00 | -0.02 *** | 0.00 | -0.04 *** | 0.00 | -0.02 *** | 0.00 |
| day × HVI class [Extreme] | -0.03 *** | 0.01 | 0.01 ** | 0.00 | -0.05 *** | 0.00 | -0.02 *** | 0.00 | -0.05 *** | 0.00 | -0.04 *** | 0.00 |
| **Random Effects** | | | | | | | | | | | | |
| $\sigma^2$ | 1.64 | | 0.82 | | 0.63 | | 0.63 | | 0.50 | | 0.54 | |
| $\tau_{00}$ | 1.67 id | | 0.39 id | | 0.27 id | | 0.19 id | | 0.13 id | | 0.31 id | |
| ICC | 0.50 | | 0.32 | | 0.30 | | 0.23 | | 0.21 | | 0.36 | |
| N | 100 id | | 100 id | | 100 id | | 100 id | | 100 id | | 100 id | |
| Observations | 61200 | | 61200 | | 61200 | | 61200 | | 61200 | | 61200 | |
| Conditional $R^2$ | 0.522 | | 0.385 | | 0.358 | | 0.347 | | 0.401 | | 0.457 | |

* *p<0.05* ** *p<0.01* *** *p<0.001*

## Appendix E Complex Contagion During Heatwave

We analyze the diffusion of heatwave-related content in our simulations to test competing theories of information contagion. Specifically, we ask whether simple contagion—diffusion through weak ties and broad exposure—is sufficient, or whether complex contagion—diffusion reinforced by cohesive, redundant networks—is necessary for uptake (Centola and Macy 2007; Guilbeault, Becker, and Centola 2018). To address this, we proceed in three stages: labeling messages, constructing interaction networks, and estimating how network features shape the likelihood of adoption.

Step 1, Labeling content. We used Llama 3.2 to classify messages generated by earlier simulations as either "relevant to heatwave" or "irrelevant to heatwave," based on a structured system prompt. Out of 45,487 total messages, 20,246 were identified as heatwave-relevant.

Step 2: Constructing networks. Interaction networks were operationalized from digital trace data by treating each unique account ID as a node (actor) and each listed friend ID as a dyadic tie. Edges were added between focal actors and alters that were also present in the valid user set. Networks were modeled as undirected, implicitly assuming symmetry of association. Temporal slices were created by grouping records by day, generating a separate graph for each period. Each daily graph contains the active nodes for that day and edges derived from same-day friend listings, yielding time-specific interaction structures. For visualization, actors mentioning the heatwave were colored red, while others were displayed in blue (Appendix E Figure 1).

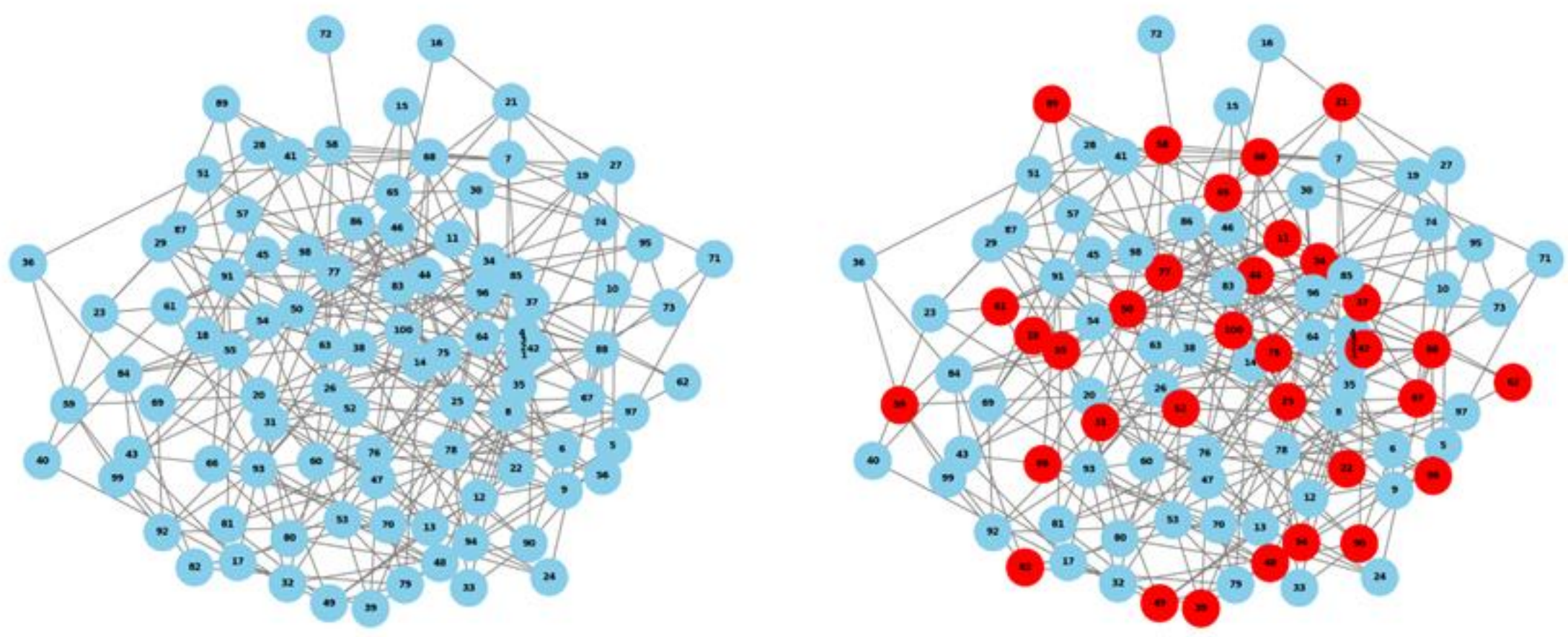

**Appendix E Figure 1. Chatting about heatwave (red nodes, day1 left and day5 right).**

Step 3: Estimating influence. To test how network embeddedness shapes discourse, we examined whether an actor discussed the heatwave on day t as a function of their network features on day t–1. Three predictors were constructed: (a) degree as a measure of exposure volume, (b) local clustering coefficient as an indicator of cohesive reinforcement, and (c) lagged peer diffusion—the number of friends who mentioned the heatwave on day t–1. A logistic regression was then fit to assess the effect of these features on next-day discourse. Methodologically, the one-day lag and the restriction to non-speakers at $t-1$ help reduce simultaneity (the classic reflection problem) inherent in concurrent models.